\documentclass[aps,prd]{revtex4}
\usepackage{epsfig,macros}
\usepackage[T1]{fontenc}
\usepackage[latin1]{inputenc}
\usepackage{graphics}

\makeatletter

\begin{document}

\begin{flushright}
  COLO-HEP-451
\end{flushright}

\title{\Large\bf Dynamical Fermions with Fat Links }
\author{Francesco Knechtli $^{\dagger}$ and Anna Hasenfratz $^{\ddagger}$}
\address{Physics Department, University of Colorado, Boulder, CO 80309 USA
}
\begin{abstract}
We present and test 
a new method for simulating dynamical fermions with fat
links. Our construction is based on the introduction of auxiliary but
dynamical gauge fields and works with any fermionic action and
can be combined with any fermionic updating. In our simulation we use
an over-relaxation step which makes it effective. For four flavors of
staggered fermions first results indicate that flavor symmetry at a lattice
spacing $a\approx0.2\fm$ is restored to a few percent. With the standard action
this amount of flavor symmetry restoration is achieved at
$a\approx0.07\fm$. We estimate that the overall computational cost is reduced by at
least a factor 10.
\end{abstract}

\maketitle
\vskip 5in
\begin{flushleft}
  December 2000
\end{flushleft} 
$^{\dagger}$ {\small e-mail: knechtli@pizero.colorado.edu} \\
$^{\ddagger}$ {\small e-mail: anna@eotvos.colorado.edu}

\vfill

\eject

\section{Introduction }

Improved actions reduce lattice artifacts and thus allow simulations at coarser
lattice spacing, larger lattice quark masses and smaller lattice volumes. The
use of improved actions in large scale numerical simulations has been increasing
steadily. For dynamical simulations a factor of two improvement in lattice spacing
can easily translate to a gain of 100 in computational cost, which usually more
than compensates for the reduction in efficiency of the algorithm. Systematic
improvement programs remove lattice artifacts perturbatively 
\cite{Symanzik:1983dc,Symanzik:1983gh,Lepage:1996ph,Alford:1998yy}
or non-perturbatively 
\cite{Hasenfratz:1994sp,DeGrand:1995ji,DeGrand:1995jk,Bietenholz:1996cy}
by adding irrelevant operators to the action. The use of fat or smeared links
is part of many of these improvement programs \cite{DeGrand:1998pr,Niedermayer:2000yx}.
Smearing the links of a lattice action does not change
the long-distance properties of the system but by smoothing out short scale
lattice vacuum fluctuations, it reduces lattice artifacts. Fat link actions by
themselves show improved scaling properties, especially in quantities most sensitive
to short distance fluctuations. 

For Wilson-type clover fermions, chiral properties show significant improvement
in fat link quenched simulations \cite{DeGrand:cloverfat}. Fattening removes
many small dislocations and that reduces the spread of the real eigenmodes of
the Dirac operator and the occurrence of exceptional configurations. The perturbation
theory for fat-link Wilson-type fermions has been worked out in \cite{Bernard:1999kc},
showing that the additive mass renormalization is small, the renormalization
factors are very close to one and the tree-level clover coefficient $c_{\rm SW}=1.0$
is expected to be close to the non-perturbative value.

Fat links have also been successfully used in overlap actions
\cite{Bietenholz:2000iy,DeGrand:2000tf}.
The improved chiral properties of fat link actions result in significantly faster
convergence in evaluating Neuberger's formula.

The use of fat links with staggered fermions improves flavor symmetry. In the staggered
fermion formulation the four components of a Dirac spinor occupy different lattice
sites and connect to different gauge fields, leading to flavor symmetry breaking.
The flavor symmetry breaking is especially evident for the pions: only one of
the pseudoscalar mesons is a true Goldstone particle, the others are massive
even at vanishing quark mass. Since flavor symmetry breaking is basically due
to the fluctuations of the gauge fields within a hypercube and is particularly
sensitive to dislocations, local smearing of the gauge links is very effective
in reducing flavor symmetry breaking.
Several quenched simulations verified
this conjecture \cite{Blum:1997uf,Lagae:1998pe,Orginos:1999cr}.
Dynamical simulations
with one level of smearing \cite{Orginos:1998ue,Karsch:2000ps} found similar
improvement.
Perturbative studies of flavor symmetry breaking also support the use
of fat links \cite{Lepage:1998vj}.

Dynamical simulation of fermion actions with fat links can be very complicated.
Even in the simplest case where the fat link is constructed as a sum of several
paths connecting the fermions, the fermionic force term will have many more
terms than with thin link action. If the fat link is projected back to SU(3)
and the fattening procedure is iterated (as proved to be most effective in quenched
simulations), direct calculation of the fermion force term becomes nearly impossible.

In this article we present a new method for simulating fat link fermion actions
with many levels of projected smearing. The basic idea is to introduce {\em an
auxiliary but dynamical gauge field} for each smearing level. These gauge fields
couple to each other by blocking kernels representing one level of smearing.
The last of the auxiliary gauge fields couple directly to the fermions just like
ordinary thin links, thus avoiding the complicated gauge force computations.
Our construction does not consider the systematic improvement of the action,
but it can be combined with any thin link fermionic action. Combining systematic
improvements of a thin link action with fat link fermions can lead to further
systematic improvement.

To motivate our choice of fat link action in Sect. 2 we study flavor
symmetry breaking with staggered fermions in the quenched approximation. We
consider valence actions with different levels of smearing and we compare the results
with the standard thin link action.
The quenched simulation suggests that with three levels
of projected fat links flavor symmetry improves to a level corresponding
to a factor of 2.5 change in lattice spacing. 

In Sect. 3 we present our general construction of a fat link fermion
action. We suggest combining different updating methods in the numerical simulation.
The simplest is to update the original and all but the last auxiliary gauge
fields using Metropolis updating. The last level auxiliary gauge field couples
to the fermions. Any fermionic updating can be used, we consider molecular dynamics
updating here. 

Unfortunately this coupled system is very rigid and evolves extremely slowly
under local updating. In Sect. 4 we discuss a global over-relaxation step that
improves the situation considerably. Over-relaxation updating based on the gauge
action is usually not very effective for fermions \cite{Hasenbusch:1998yb}. The
situation here is quite different. Since the fermions couple to a several times
smeared smooth gauge field, we find that one can update up to $(0.3\fm)^4$
part of the lattice and still have a large enough acceptance rate to make the
algorithm efficient. 

In Sect. 4 we specify the action for four flavors of staggered
fermions and discuss the over-relaxation update in detail. One iteration of
our algorithm is composed of 100 over-relaxation, one Metropolis and one
molecular dynamics steps. This combined updating is
about 15-20 times slower than a molecular dynamics thin link updating and has
about the same autocorrelation times. Considering that we gain well over a factor
of 100 from the improved scaling properties, this cost is acceptable. 
Our first results with this algorithm 
confirms the quenched results for flavor symmetry breaking. We find
that flavor symmetry violations are reduced to a
few percent at a lattice spacing $a\approx0.20\fm$.

In Sect. 5 we summarize our results and discuss the future directions.

\section{Fat Link Actions and Quenched Spectroscopy }

In this section we investigate the spectrum of fat link staggered 
actions in the quenched
approximation. 
Previous extensive studies \cite{Orginos:1999cr}
have demonstrated the improvement in the restoration of flavor symmetry due to
the smearing of the gauge links.
Our goal here is to motivate the parameters of our dynamical fat link action.
\begin{figure}[tb]
\hspace{0cm}
\vspace{-1.0cm}
\centerline{\psfig{file=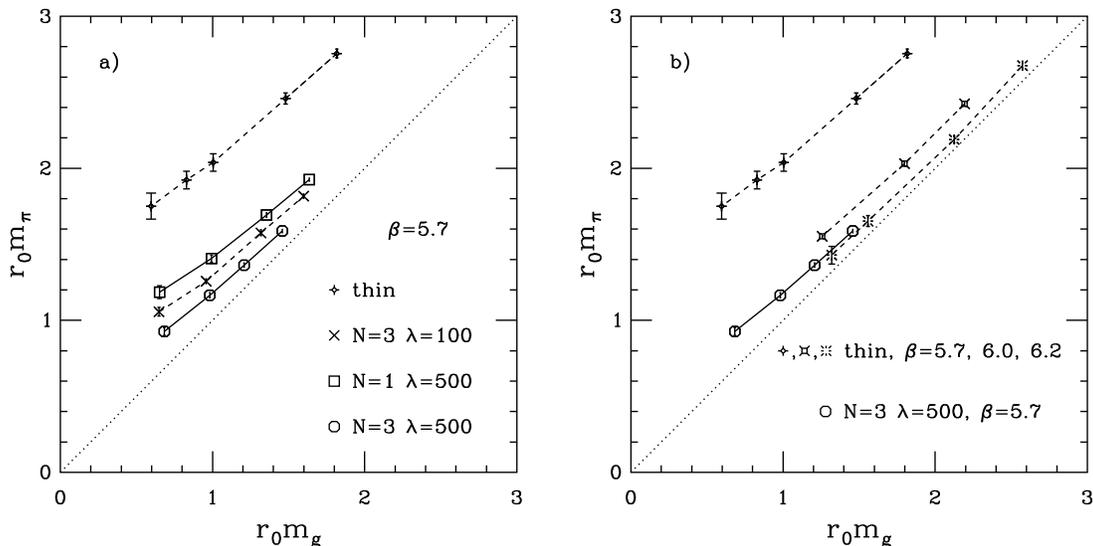,width=14.5cm}}
\vspace{0.2cm}
\caption{The mass $m_{\pi}$ of the lightest non-Goldstone pion as the 
  function of the mass $m_g$ of the Goldstone pion
  in the quenched approximation. Fig. 1a)
  shows results for $\beta=5.7$ using both a thin link
  and different fat link valence actions. 
  In Fig. 1b) results obtained with our best fat link action on 
  $\beta=5.7$ background are compared with thin link action  results on $\beta=5.7$, 
  $\beta=6.0$  and $\beta=6.2$ background configurations. The lattice spacing
  changes by a factor of 2.5 between $\beta=5.7$ and $\beta=6.2$.
  \label{f_pi_vs_G}}
\end{figure}

We fatten the gauge links using 
APE smearing \cite{smear:ape}: the smeared or fat link $Q$ is
constructed from the thin link $U$ as
\bes\label{fat}
 Q_{i,\mu} & = & (1-\alpha)U_{i,\mu} + \frac{\alpha}{6}\Sigma_{i,\mu}(U) \,,
\ees
where $\Sigma_{i,\mu}(U)$ is the sum over the six staples around the link
$U_{i,\mu}$. We use the index $i$ to label the lattice sites and 
the index $\mu$ to label the four space-time directions.
The smearing procedure \eq{fat} can be iterated if the fat link $Q$ is projected to SU(3)
\bes\label{proj}
 W_{i,\mu}=Proj_{\SUthree}\{Q_{i,\mu}\}.
\ees
 The $n$-th level fat link is given by
\bes\label{fatn}
 Q^{(n)}_{i,\mu} & = & (1-\alpha)W^{(n-1)}_{i,\mu} +
 \frac{\alpha}{6}\Sigma_{i,\mu}(W^{(n-1)})  \,,
\ees
where $\Sigma_{i,\mu}(W^{(n-1)})$ is the sum of staples around $W_{i,\mu}^{(n-1)}$,
the $(n-1)$-th level fat link projected onto SU(3) ($W^{(0)}\equiv U$).
In the following 
we label by $N$ the number of smearing iterations or levels.
Perturbative arguments \cite{Bernard:1999kc} show that for values of the
smearing parameter $0\leq\alpha\leq0.75$ the smearing orders the gauge configuration
suppressing small scale fluctuations. If $\alpha>0.75$ the smearing eventually
disorders. In the following we choose, somewhat arbitrarily, $\alpha=0.70$.

We consider two different SU(3) projections :
 a {\em deterministic} projection $W_{{\rm max},i,\mu}$ defined by
\bes\label{maxproj}
 {\rm Re}\Tr(\,W_{{\rm max},i,\mu}\,Q_{i,\mu}^{\dagger}\,) & = & 
 \max_{W\in{\rm SU(3)}}\,{\rm Re}\Tr(\,W\,Q_{i,\mu}^{\dagger}\,)
\ees
and a {\em probabilistic} projection $W^{\lambda}_{i,\mu}$,
where $W^{\lambda}_{i,\mu}$ is chosen  
according to the probability distribution
\bes\label{probproj}
 P(W) & \propto & \exp[\frac{\lambda}{3}{\rm Re}\Tr(WQ_{i,\mu}^{\dagger})]
\ees
with projection parameter $\lambda$. For $\lambda=\infty$, \eq{probproj} is equivalent
to \eq{maxproj}.
\begin{figure}[tb]
\hspace{0cm}
\vspace{-1.0cm}
\centerline{\psfig{file=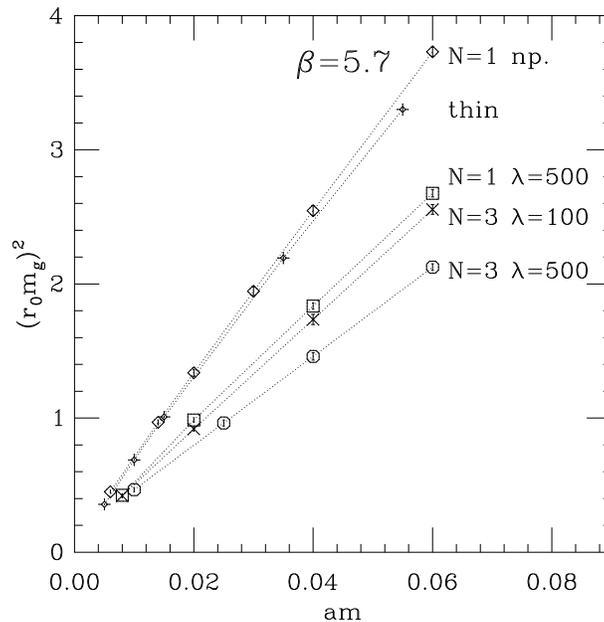,width=8cm}}
\vspace{0.2cm}
\caption{The mass renormalization in the quenched
  approximation at $\beta=5.7$ for the thin link and different fat
  link valence actions. In addition to the actions used in \fig{f_pi_vs_G}, we
  show the results for $N=1$ level of non-projected (np.) APE smearing.  
 \label{f_G_vs_mq}}
\end{figure}

To illustrate the effect of fat links on flavor symmetry restoration we
calculated the pion spectrum on a set of $8^3×24$, $\beta=5.7$ ($a=0.17$fm) 
quenched lattices.
In \fig{f_pi_vs_G}a) we plot the masses of the (would-be) Goldstone
pion $\pi_g$ and the lightest non-Goldstone pion,
$\pi_{i5}$, corresponding to the representation
$\gamma_5\otimes\gamma_i\gamma_5$. We label the representation
of the states following \cite{Kluberg-Stern:1983dg,Gupta:1991mr} by
$\Gamma_S\otimes\Gamma_F$, where $\Gamma_S$ labels the spin and $\Gamma_F$
labels the SU(4) flavor.
The pion masses are expressed in units of the Sommer
scale $\rnod$ ($r_0=2.87a$ at $\beta=5.7$) \cite{Sommer:1994ce}.
In addition to the thin link we consider three fat link valence actions:
$N=1$ and $N=3$ levels of smearing with projection parameter $\lambda=500$
and  $N=3$ levels of smearing with
projection parameter $\lambda=100$, all with $\alpha=0.7$.
We observe considerable improvement from $N=1$ to $N=3$. 
Also,  $\lambda=100$
for the projection parameter is clearly not as effective as $\lambda=500$.
Increasing the number of blocking levels N or the projection parameter $\lambda$ further
does not improve the situation substantially but increases the computational effort
considerably. In the rest of this paper we will consider the fat link action
corresponding to N=3 levels of blocking with projection parameter $\lambda=500$
and APE parameter $\alpha=0.7$. While these parameters are not unique, they seem
to be close to optimal.

To get a feel for the amount of improvement one can achieve with this 
fat link actions, in \fig{f_pi_vs_G}b) we compare the $N=3$, $\lambda=500$ 
action with thin link
data at $\beta=5.7$, 6.0 and 6.2. The data for the last two $\beta$ 
values are from ref. \cite{Gupta:1991mr} and correspond to lattice spacings 
$a=0.095\fm$ and $0.069\fm$
and Sommer parameters $r_0=5.26a$ and $7.25a$, respectively \cite{Guagnelli:1998ud}.
The $N=3$ , $\lambda=500$ smearing reduces flavor symmetry violations on the
$\beta=5.7$ configurations to the level obtained at $\beta=6.2$ with thin link
action, a factor of 2.5 improvement in lattice spacing.
The computational cost of full
QCD grows at least as $1/a^6$ \cite{Lepage:1996ph} and for certain quantities
even as  
$1/a^{10}$ \cite{Niedermayer:1997eb}.
A factor 2.5 in the lattice spacing means a factor $200-10^4$
in the computational costs.

Finally, in \fig{f_G_vs_mq} we plot the square of the 
Goldstone-pion mass as a
function of the bare quark mass. 
All measurements are on the  $\beta=5.7$ gauge configurations using
thin link and different fat link
actions. In addition to the actions considered in \fig{f_pi_vs_G}, we show the
results for $N=1$ level of non-projected APE smearing.
Fat link perturbation theory predicts
that the mass renormalization constant becomes perturbative as N increases.
We see that the mass
renormalization after one level of non-projected smearing is almost the same
as without smearing. Increasing the smearing level and the value of $\lambda$
indeed reduces 
the multiplicative mass renormalization factor.

\section{Dynamical Fermions with Fat Links }

The quenched results 
show that smearing the gauge links 
considerably improves the flavor symmetry of staggered
fermions. There is evidence that these results carry over to dynamical
simulations \cite{Orginos:1998ue} where one level of smearing is implemented.
Because of the gauge force computations in the molecular dynamics equations of
motion, it is very complicated
to simulate fermions with many levels of smearing, 
even impossible if projection of the gauge
links onto SU(3) is made after each smearing step.

Our method to overcome this difficulty
is to introduce of a set of {\em auxiliary but dynamical gauge fields}
which couple to each other in the action 
by blocking kernels representing one level of smearing.
The last level auxiliary gauge field couples directly to the
fermions in the same way the usual thin links do. The problem of
the computations of the gauge force is transferred to the
gauge sector where it can be solved.
This construction can be used for any fermion action
which can be simulated with ``thin'' links.

Let us start considering fermions coupled to fat links constructed with one
level of smearing from the thin links $U_{i,\mu}$. We introduce a dynamical
auxiliary gauge field $V$ and we define the action
\bes\label{action2}
 S & = & -\frac{\beta}{3}\sum_{\rm p}{\rm Re}\Tr(U_{\rm p})
 -\frac{\lambda}{3}\sum_{i,\mu}{\rm Re}\Tr(V_{i,\mu}W^{\dagger}_{{\rm max},i,\mu}(U))
 -{\rm tr}\ln[M^{\dagger}(V)M(V)] \,.
\ees
Here $p$ labels the plaquettes $U_p$, $W_{\rm max}(U)$ is the
SU(3) projection \eq{maxproj} of the fat link given in \eq{fat} and $M(V)$ is
the fermion matrix. The blocking parameter $\lambda$ constrains the
auxiliary gauge links $V_{i,\mu}$ to be close to the projected fat links
$W_{{\rm max},i,\mu}(U)$. Fluctuations of the field $V$ are proportional to
$1/\lambda$. This is a dynamical realization of the projection \eq{probproj}.
In \eq{action2}, $\Tr$ means the trace over SU(3) color whereas $\tr$
means the overall trace over space-time indices $i$, directions $\mu$, spin and
color.

For many levels of smearing, we introduce a set of
dynamical auxiliary gauge fields $W^{(1)},W^{(2)},\ldots W^{(N)}\equiv V$, one for each level of smearing.
The action \eq{action2} is generalized to
\bes\label{actionmany}
 S & = & -\frac{\beta}{3}\sum_{\rm p}{\rm Re}\Tr(U_{\rm p})
 -\frac{\lambda}{3}\sum_{i,\mu}{\rm Re}\Tr(W_{i,\mu}^{(1)}W^{\dagger}_{{\rm max},i,\mu}(U))
 \nonumber \\
& & -\frac{\lambda}{3}\sum_{i,\mu}{\rm Re}\Tr(W_{i,\mu}^{(2)}W^{\dagger}_{{\rm max},i,\mu}(W^{(1)})) 
\;\ldots\;  -\frac{\lambda}{3}\sum_{i,\mu}{\rm Re}\Tr(V_{i,\mu}W^{\dagger}_{{\rm max},i,\mu}(W^{(N-1)})) 
 \nonumber \\
& & -{\rm tr}\ln[M^{\dagger}(V)M(V)]\,.
\ees
This action is in the same universality class as the original thin link
action. If the blocking parameter is $\lambda=\infty$, the auxiliary gauge fields can
be integrated out and the resulting action is a plaquette gauge action while
the fermions couple to deterministically projected fat links. If $\lambda\neq\infty$, the
integration of the auxiliary gauge fields will introduce additional gauge
terms. These new terms depend on the thin link variables in a complicated way,
but they are all local terms containing only a finite number of link
variables. Thus, these terms do not change the universality class of the action.

The updating of this system can, in principle, be done
by a sequence of Metropolis updatings for the fields
$U,W^{(1)},\ldots W^{(N-1)}$ and by the standard updating for the fermion matrix $M$,
with the only difference that now the last auxiliary gauge field $V$ enters
the fermion matrix.

The problem of this basic algorithm is that
the system with large $\lambda$ is very rigid and evolves very slowly.
To cure this problem,
we use a hybrid over-relaxation algorithm
\cite{Gupta:1988pf,Apostolakis:1991km,Hasenbusch:1992tx,Wolff:1992ze,Creutz:1987xi,Brown:1987rr,Decker:1990hp,Gupta:1988yw,Booth:1992kk,Wolff:1992ri}
with
\begin{itemize}
\item[$\star$] Metropolis updating for the gauge fields $U,W^{(1)},\ldots W^{(N-1)}$,
\item[$\star$] standard algorithm for the fermion matrix $M(V)$ and
\item[$\star$] {\em global over-relaxation} for all the gauge fields $U,W^{(1)},\ldots V$
\end{itemize}
In the next section we discuss how an over-relaxation can be implemented.
It plays a key role in reducing the autocorrelation times of the 
Metropolis update. Note that the over-relaxation update is effective only
because the fermions couple to the smooth fat links.

\section{Over-relaxation with Fat Links }

Our over-relaxation update 
is based on the usual over-relaxation reflection step used in 
pure gauge systems that leaves the gauge action invariant
\cite{Creutz:1987xi,Brown:1987rr}. 
We reflect the thin links just like in the standard over-relaxation algorithm
by sweeping in a given order through some part (or all) of the lattice
\bes\label{urefl}
U & \to & U^{\prime} \,.
\ees
These changes are followed by a transformation of the fat links 
\bes
 W^{(1)\prime} & = & W^{(1)}W^{\dagger}_{\rm max}(U)W_{\rm max}(U^{\prime})
 \label{refl1} \,, \\
 W^{(2)\prime} & = & W^{(2)}W^{\dagger}_{\rm max}(W^{(1)})W_{\rm max}(W^{(1)\prime})
 \label{refl2} \,, \\
\ldots & & \nonumber \\
 V^{\prime} & = & VW^{\dagger}_{\rm max}(W^{(N-1)})W_{\rm max}(W^{(N-1)\prime})
 \label{reflN} \,.
\ees
All links for a given
level are reflected and then
the next level gauge field is changed. The reflections must be performed in the
order
given by \eq{urefl}-\eq{reflN}. This transformation leaves the gauge part of the action 
invariant, but the fermionic part will change and a Metropolis accept/reject 
step must be performed.
In order for this updating to satisfy detailed balance, the probability $P$
for changing the gauge field configuration $\{U,W^{(1)},\ldots V\}$ to $\{U^{\prime},W^{(1)\prime},\ldots V^{\prime}\}$ 
has to be equal to the probability for the reversed change, i.e.
\bes\label{symm}
 P(\{U,W^{(1)},\ldots V\}\to\{U^{\prime},W^{(1)\prime},\ldots V^{\prime}\}) & = &
 P(\{U^{\prime},W^{(1)\prime},\ldots V^{\prime}\}\to\{U,W^{(1)},\ldots V\}) \,.
\ees
This can be achieved by choosing with probability $1/2$ either a given
sequence of the thin link reflections \eq{urefl} or with equal probability the
reversed sequence \cite{Hasenbusch:1998yb}. The sequence has to be reversed
with respect to the direction and location of the thin links and 
with respect to the index of the SU(2) subgroup. The thin link reflections are
then followed by the local reflections
\eq{refl1}-\eq{reflN} for the auxiliary gauge links.
For a given level of auxiliary gauge field the reflections
are independent of the order with which we sweep through the lattice and they are
symmetric under exchange of primed and unprimed quantities. If we start from
a gauge field configuration $\{U,W^{(1)},\ldots V\}$ and apply
\eq{urefl}-\eq{reflN} twice with the sequence of thin link reflections
reversed, we come back to the original configuration.
This over-relaxation algorithm satisfies detailed balance and is
a legal update of the system.
To achieve ergodicity though, one must
still perform some updatings with the basic algorithm.

For the Metropolis accept-reject step following the changes \eq{urefl}-\eq{reflN}
we have to calculate the action, i.e. we need an explicit form 
of the determinant to evaluate. 
For four flavors of staggered and two flavors of Wilson fermions this 
is trivially realized.
In terms of pseudofermion fields $\Phi$ the action  
\eq{actionmany} is
\bes\label{actionmany2}
 S & = & S_g(U,W^{(1)},\ldots V) + \Phi^{\dagger}[M^{\dagger}(V)M(V)]^{-1}\Phi  \,,
\ees
where  M(V) represents the fermion matrix.
The equilibrium probability distribution of the gauge fields
$U,W^{(1)},\ldots V$ is given by
\bes\label{eqprob0}
 P_{\rm eq}(U,W^{(1)},\ldots V) & 
\propto & \rme^{-S_g(U,W^{(1)},\ldots V)}\,\det(M^{\dagger}(V)M(V)) \,.
\ees
Since the pure gauge part of the action is invariant under the over-relaxation
move of the system, the acceptance probability is 
\bes\label{accept0}
P_{\rm acc}(V^{\prime},V) & = & \min\left\{1,\frac{\det(M^{\dagger}(V^{\prime})M(V^{\prime}))}{\det(M^{\dagger}(V)M(V))}\right\}
\,.
\ees
$P_{\rm acc}$ depends only on the last level auxiliary gauge links.
The goal is to efficiently compute the acceptance probability.

Instead of calculating the determinant in \eq{eqprob0} we use a stochastic 
estimator to approximate $P_{\rm acc}$\cite{Grady:1985fs,creutz_algo}
\bes\label{accept1}
 P_{\rm acc}^{\prime}(V^{\prime},V) & = & \min\Big\{1,\rme^{
 \xi^{\dagger}[M^{\dagger}(V^{\prime})M(V^{\prime})-M^{\dagger}(V)M(V)]
\xi}\Big\} \,,
\ees
where the  vector $\xi$ is
generated according to the distribution
\bes\label{xsidistr0}
 P(\xi) & \propto & \rme^{-\xi^{\dagger}M^{\dagger}(V^{\prime})M(V^{\prime})
\xi} \,.
\ees
After configuration averaging, 
this procedure satisfies the detailed balance condition \cite{creutz_algo}.

If the gauge fields $V$ and $V^{\prime}$ are very different, 
the acceptance rate from \eq{accept1} will be small even when 
the fermionic determinants are actually very close.
This is because of the large fluctuations of the stochastic estimator.
To improve the acceptance rate, we 
attempt to remove the most ultraviolet
part of the fermionic matrix and include it explicitly as an effective gauge
action. The resulting reduced matrix gives a much higher 
acceptance in \eq{accept1}. 

For heavy fermions
the fermion determinant gives rise to an effective
plaquette term for the gauge field \cite{Hasenfratz:1994az}.
Even for small quark masses the ultraviolet part
of the fermion determinant can be well approximated by an
effective loop action involving only small Wilson loops
\cite{Duncan:1998gq,Duncan:1999xh,deForcrand:1998sv}.
These observations suggest to remove the plaquette term from
the fermion matrix by introducing a reduced matrix $M_r$ as
\bes
 M(V) & = & M_r(V)A(V) \qquad \mbox{with} \label{fermionma} \\
 A(V) & = & \rme^{\alpha_4D^4(V)+\alpha_2D^2(V)} \,, \label{mata}
\ees
where $D$ is the kinetic part of the fermion matrix.
In terms of $M_r$ the fermion determinant becomes
\bes
 \det(M^{\dagger}(V)M(V)) & = & \det(M_r^{\dagger}(V)M_r(V))\,\rme^{-S_{\rm eff}(V)}
 \,, \label{detnew}\\
 S_{\rm eff}(V) & = & -2\alpha_4{\rm Re}\,{\rm tr}[D^4(V)]-2\alpha_2{\rm Re}\,{\rm tr}[D^2(V)]
 \,. \label{seff}
\ees
The effective 
action $S_{\rm eff}(V)$  can be evaluated explicitly. In general it 
is the sum of a  plaquette term
coming from ${\rm tr}[D^4(V)]$ and a constant from ${\rm tr}[D^2(V)]$.
The real parameters $\alpha_2$ and $\alpha_4$ are free and can be optimized.
In a different context, a decomposition of
the fermion matrix like \eq{fermionma} has been proposed in
\cite{Hasenbusch:1998yb} motivated by the hopping parameter expansion.

In terms of $S_{\rm eff}(V)$ and $M_r(V)$ the acceptance probability is
\bes\label{accept0new}
 P_{\rm acc}(V^{\prime},V) & = & \min\left\{1,\frac{\rme^{-S_{\rm eff}(V^{\prime})}\det(M_r^{\dagger}(V^{\prime})M_r(V^{\prime}))}
       {\rme^{-S_{\rm eff}(V)}\det(M_r^{\dagger}(V)M_r(V))}\right\} \,.
\ees
It can be approximated by generating a vector $\xi$ according to the distribution
\bes\label{xsidistr}
 P(\xi) & \propto & \rme^{-\xi^{\dagger}M^{\dagger}_r(V^{\prime})M_r(V^{\prime})\xi}
\ees
and \eq{accept0new} becomes
\bes\label{accept}
 P_{\rm acc}^{\prime\prime}(V^{\prime},V) & = & \min\Big\{1,\rme^{S_{\rm eff}(V)-S_{\rm eff}(V^{\prime})
 +\xi^{\dagger}[M^{\dagger}_r(V^{\prime})M_r(V^{\prime})-M^{\dagger}_r(V)M_r(V)]\xi}\Big\} \,.
\ees
In practice, we start by generating a random Gaussian source $R$
according to the probability distribution
\bes\label{rangauss}
 P(R) & \propto & \rme^{-R^{\dagger}R}
\ees
from which we form the vectors
\bes\label{phixp}
 \Phi^{\prime} & = & M^{\dagger}(V^{\prime})R \,,\\
 X^{\prime} & = & [M^{\dagger}(V^{\prime})M(V^{\prime})]^{-1}\Phi^{\prime} \,.
\ees
The vector $\xi$ in \eq{xsidistr} is then given by
$\xi=A(V^{\prime})X^{\prime}$
and we can write the fermionic terms in \eq{accept} as
\bes
 \xi^{\dagger}M^{\dagger}_r(V^{\prime})M_r(V^{\prime})\xi & = & \Phi^{\prime\dagger}X^{\prime} \,,\\
 \xi^{\dagger}M^{\dagger}_r(V)M_r(V))\xi & = &
 X^{\prime\dagger}A^{\dagger}(V^{\prime})A^{\dagger}(V)^{-1}M^{\dagger}(V)M(V)A(V)^{-1}A(V^{\prime})X^{\prime} \,.
\ees
The  acceptance probability  strongly depends on 
the parameters $\alpha_2$ and $\alpha_4$.
The optimal value can be found numerically.
We will discuss our
choice in the next section.

This procedure is not effective if the fermion matrix depends on thin
links, because the fluctuations in the stochastic estimator 
are too
large even after the removal of the $D^4$ and $D^2$ terms. 
When the links in the fermion matrix are smeared, these
fluctuations are constrained. This is the key feature which
makes the over-relaxation effective with fat links.

\section{Performance of the Algorithm }

For testing our  fat link action
we decided to simulate $N_f=4$ flavors of staggered fermions.
The fermion matrix is given by
\bes\label{staggered}
 M(V)_{i,j} & = &
 2m\delta_{i,j}+\sum_{\mu}\eta_{i,\mu}(V_{i,\mu}\delta_{i,j-\muh}-V^{\dagger}_{i-\muh,\mu}\delta_{i,j+\muh}) \,,
\ees
where $\eta_{i,\mu}$ are the staggered phases.
We impose anti-periodic boundary conditions in the time
direction.
The pseudofermion field $\Phi$ and the matrix $M^{\dagger}(V)M(V)$
are  restricted to the even sites of the lattice \cite{Martin:1985yn}.
The matrix $D$ used in \eq{mata} is given by
\bes\label{dstag}
 D_{i,j} & = &
 \sum_{\mu}\eta_{i,\mu}(V_{i,\mu}\delta_{i,j-\muh}-V^{\dagger}_{i-\muh,\mu}\delta_{i,j+\muh}) \,.
\ees
The traces in $S_{\rm eff}(V)$ in \eq{seff} are computed by summing over the even
sites only and give
\bes
 {\rm tr}[D^4(V)] & = & 24\Omega[3-\frac{1}{6\Omega}\sum_{\rm p}{\rm Re}\,\Tr(V_{\rm  p})]+108\Omega
                   -4\delta_{N_t,4}\sum_{i,t=0}{\rm Re}\,\Tr(P_i) \,, \label{trd4}\\
 {\rm tr}[D^2(V)] & = & -12\Omega \,.
\ees
$\Omega$
denotes the total number of lattice points and we assume that there are more than
4 sites in each space-like direction.
For $N_t=4$ sites in the time-like
direction there is an extra contribution to ${\rm tr}[D^4(V)]$ coming from the
Polyakov lines $P_i$ starting at location $i,t=0$.
The minus sign of the Polyakov line in \eq{trd4}
is due to the anti-periodic boundary conditions in time direction.

As we described in Sect. 2
we choose the number of auxiliary gauge fields and the smearing parameters to be
\bes\label{params}
 N=3 \,, & \alpha=0.7 \,, & \lambda=500 \,.
\ees
\begin{figure}[tb]
\hspace{0cm}
\vspace{-1.0cm}
\centerline{\psfig{file=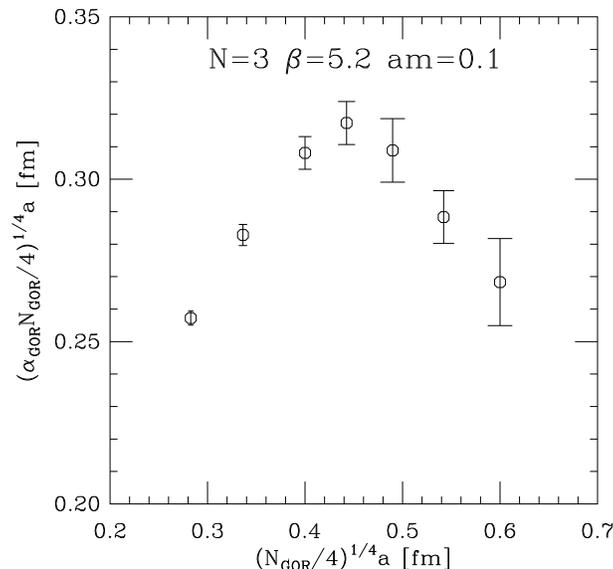,width=8cm}}
\vspace{0.2cm}
\caption{The effective physical volume
  where the $U$ links change in one
  global over-relaxation (GOR)
  updating step as function of the physical volume ``touched''. The results
  are obtained on $8^3×24$ lattices.
  \label{f_accept}}
\end{figure}

The global over-relaxation
(GOR) described in Sect. 4 is essential here as it
considerably reduces the autocorrelations of the otherwise 
very rigid system.  The GOR leaves the pure gauge action of our system
invariant but is subjected to an accept/reject step which accounts for the
ratio of fermion determinants, see \eq{accept}. Since the fermions couple directly
to the last level dynamical fat links, the acceptance rate $\alpha_{\rm GOR}$ is
large enough to make the algorithm effective.

The parameters $\alpha_2$ and $\alpha_4$ entering \eq{mata} are chosen to 
maximize the acceptance rate $\alpha_{\rm GOR}$. We use
\bes\label{alphamata}
 \alpha_2=-0.18 &\,,\, & \alpha_4=-0.006.
\ees
Setting $\alpha_2=\alpha_4=0$, i.e. computing the ratio of fermion determinants
without decomposing the fermion matrix according to \eq{fermionma} and
\eq{mata} gives a value for $\alpha_{\rm GOR}$ which is
a factor 10 smaller than what we achieve with our choice. Moreover, keeping
$\alpha_4=0$ and varying only $\alpha_2$ gives significantly lower
 $\alpha_{\rm GOR}$.
The choice of \eq{alphamata}
is not unique,
we identified in the $\alpha_2-\alpha_4$ parameter space a band-like region in
which $\alpha_{\rm GOR}$ reaches its maximal value.

Even with our improved GOR
it is not possible to change simultaneously all the $U$ links, 
the effectiveness of the algorithm would be very low. Instead
we  choose a random block  of $U$ links
containing $(N_{\rm GOR}/4)$ sites and 
reflect only the links within this block. 
These changes propagate
more and more through the lattice as we consider the cascade of reflections
\eq{urefl}-\eq{reflN}, as 19 fat links have to be changed by changing one link.
In \fig{f_accept} we plot the volume of the lattice in which the $U$ links
are effectively updated (actually the fourth root of it),
i.e. $(\alpha_{\rm GOR}N_{\rm GOR}/4)^{1/4}a$ as function of the physical volume
of the ``touched'' $U$ links $(N_{\rm GOR}/4)^{1/4}a$. These results are
obtained on $8^3×24$, $\beta=5.2$ and $am=0.1$ lattices.
The lattice spacing from the string tension can be estimated as
$a\approx0.2\fm$ and the correlation length as $\rnod m_g\approx1.7$.
We see in \fig{f_accept} that there is a maximal physical volume
\bes\label{physvolacc}
 V_{\rm GOR} & \approx & (0.3\fm)^4
\ees
which can be updated with a reasonable acceptance rate.
The actual value of $\alpha_{\rm GOR}$ depends on the number of ``touched'' links
$N_{\rm GOR}$. The broad maximum in \fig{f_accept} corresponds to
$\alpha_{\rm GOR}=35\%\;(N_{\rm GOR}=64)$,
$\alpha_{\rm GOR}=26\%\;(N_{\rm GOR}=96)$ and
$\alpha_{\rm GOR}=16\%\;(N_{\rm GOR}=144)$ on the above lattices.
\begin{table}
 \centerline{
 \begin{tabular}{|l|c|c|c||c||c|} \hline
  links & 100 $×$ GOR & 1 $×$MET & 1 $×$ HMC & total & CG iterations
  \\ \hline\hline
  thin & - & - & 1 & 1 & 133 \\ \hline
  fat & 11.0 & 3.5 & 0.5 & 15.0 & 61 \\ \hline
 \end{tabular}}
\caption{Timings for simulation of ordinary thin
  link HMC algorithm compared to simulation of
  our algorithm with $N=3$ auxiliary gauge fields. 
  The dynamical lattices are $8^3×24$ and
  the lattice spacings and physical quark masses of
  the two simulations are approximately matched.
  The time unit is one updating step (consisting of
  one HMC trajectory) of the ordinary thin link algorithm.
  The last two columns give the total time costs and the average
  number of conjugate gradient (CG) iterations needed per inversion of $M^{\dagger}M$. 
  \label{t_times}}
\end{table}

We observed that
the value \eq{physvolacc} scales with the lattice spacing.
On lattices with smaller lattice spacing more links can be updated at the same
time, i.e. the physical volume of the updated region remains fixed.
On larger physical volumes, on the other hand, one would have to increase the
number of GOR steps to achieve the same efficiency.
In our finite-temperature
runs on $8^3×4$ lattices \cite{coloprep} we observed that the effectiveness
of the algorithm increases in the deconfined phase. Somewhat surprisingly, the
effectiveness also increases with decreasing quark mass. This fact together
with the faster convergence of the conjugate gradient (see \tab{t_times})
allows simulations at quark masses that are not practically possible with the
standard thin link action.
\begin{figure}[tb]
\hspace{0cm}
\vspace{-1.0cm}
\centerline{\psfig{file=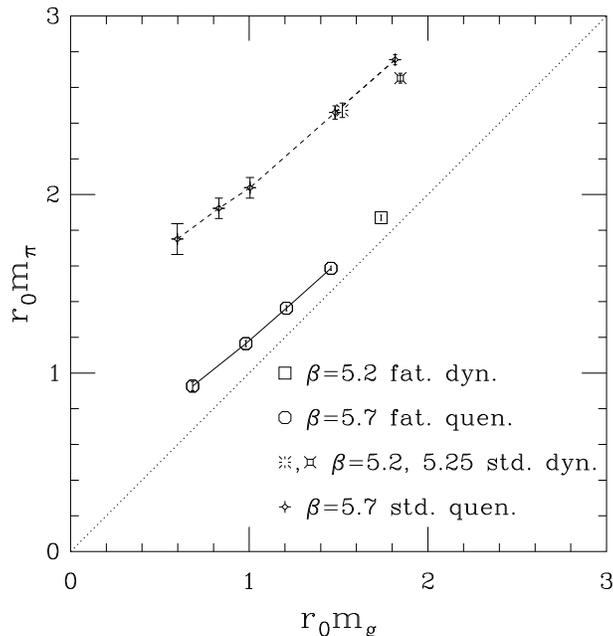,width=8cm}}
\vspace{0.2cm}
\caption{Flavor symmetry violation for the dynamical runs
  with our fat link action (fat. dyn.) and with the standard thin link 
  staggered action (std. dyn.).
  The mass $m_{\pi}$ of the lightest non-Goldstone pion is plotted as function of
  the mass $m_g$ of the Goldstone pion (in units of $\rnod$).
  The lattice spacings and the correlation lengths $\rnod m_g$
  are approximately matched. For comparison we plot the
  quenched results from \fig{f_pi_vs_G} obtained with the corresponding
  valence actions (fat. quen. and std. quen.). 
  \label{f_pi_vs_G_dyn}}
\end{figure}

In simulating the $8^3×24\approx20{\fm}^4$ lattices, we 
found it effective to follow each Metropolis and
HMC by 100 GOR updating steps, each reflecting about $(0.3\fm)^4$ section of
the thin link lattices. For timing our algorithm 
we consider the computer time necessary for  
100  GOR steps,
one Metropolis (MET) step for each gauge field $U,W^{(1)},W^{(2)}$ 
and one HMC trajectory for the $V$ field with
step size $\Delta t=0.015$ and $N_{\rm traj}=30$ steps. 
We compare this to the time of one HMC trajectory with
step size $\Delta t=0.015$ and $N_{\rm traj}=30$ steps
for the thin link action \cite{Gottlieb:1987mq,Duane:1987de}
with parameters $\beta=5.25$ and $am=0.06$.
With these parameters the lattice spacings and the physical quark masses of
the two actions are approximately matched.
To compare the times for one updating iteration
is fair because the autocorrelations for simple observables like the
plaquette or the chiral condensate $\bar{\psi}\psi$
are observed to be the same for the two algorithms in these time units.

In \tab{t_times} we show the results of the timing comparison.
We use one iteration of
ordinary thin link HMC as time unit. One iteration of
our algorithm costs a factor 15 more but, as pointed out in
Sect. 2, we can effectively gain a factor $10^2-10^4$ in computer time
due to improved scaling.
With fat links,
there is also a considerable reduction (a factor 2)
in the number of conjugate gradient (CG)
iterations needed for the inversion of $M^{\dagger}M$, as shown in the last
column of \tab{t_times}.

Finally we consider flavor symmetry restoration on the dynamical fat link
lattices.
\fig{f_pi_vs_G_dyn} shows the first results
obtained with our fat link action on $8^3×24$ lattices with
parameters $\beta=5.2$, $am=0.1$ (square). The lattice spacing is $a\approx0.2\fm$ and
the Goldstone pion to rho mass ratio is $m_g/m_{\rho}=0.71(1)$, indicating a
correlation length $\rnod m_g\approx1.7$. This point agrees very well
with the quenched results (octagons) taken from \fig{f_pi_vs_G},
i.e. flavor symmetry violation is reduced to a few percent. 
To show the improvement due to the smearing of the gauge links, we also ran
the standard thin link staggered action
with two sets of parameters,
$\beta=5.25$, $am=0.06$ (fancy square) and $\beta=5.2$, $am=0.06$ (burst). The
lattice spacings from the string tension are approximately $a\approx0.18\fm$ and
$a\approx0.22\fm$, and the correlation lengths are $\rnod m_g\approx1.8$ and $\rnod
m_g\approx1.5$, respectively.
The flavor symmetry violations on the thin link dynamical lattices agree
with the quenched predictions (fancy diamonds).
Our fat link dynamical action has only about 6\% flavor symmetry violation
compared to 60\% of thin link actions at comparable lattice spacings.

\section{Conclusions }

We presented a new method for simulating dynamical fermions with fat links
constructed through many levels of projected smearing. For each level of
smearing we introduce an auxiliary but dynamical gauge field and these gauge
fields are connected to each other by blocking kernels representing one level
of smearing. Since the last of the auxiliary fields couple in the standard way
to the fermions, our construction can be used with any known fermionic update.
We discussed the simulation of our system which includes
an over-relaxation updating. The fat links entering the fermion
matrix make the over-relaxation effective and this is the key feature of our
algorithm.

At this time our algorithm is running on scalar machines and the over-relaxation
step is worked out for four flavors of staggered fermions. The results for flavor
symmetry restoration confirm the quenched results, that is, a factor 2.5 in the
lattice spacing can be gained. Taking into account that our algorithm is about
15-20 times slower than the standard one, this gives an overall gain of at
least a factor 10 in computational costs.

We have used this algorithm to study the finite temperature phase transition
of four flavors of staggered fermions at $N_t=4$ \cite{coloprep}. We observe a
qualitative difference compared to thin link simulations. The strongly first
order phase transition is washed away, we observe a very broad crossover
instead. We believe this difference is due to the improved flavor symmetry. In
our simulations we have 15 relatively light pions compared to the single
Goldstone particle of thin link simulations.

We are parallelizing the code for large scale simulations \cite{coloprep2}.
This requires a 32-checkerboard structure but it is not more complicated than
parallelizing a Symanzik improved gauge action.
The over-relaxation generalizes easily for
two flavors of Wilson fermions. To generalize it for two flavors of staggered fermions
we need an explicit realization of the action. That can be done by approximating
the square root of the fermionic determinant with a polynomial form
\cite{Takaishi:2000bt,Montvay:1996ea}.
Our preliminary study shows that it can be done efficiently.

{\bf Acknowledgment.}
We are indebted to the MILC Collaboration for the use of their
code for standard staggered fermions.
We thank Thomas DeGrand for many useful discussions and M. Hasenbusch,
F. Niedermayer and U. Wolff for their important suggestions.
This work was supported by the U.S. Department of Energy.


\begin{thebibliography}{10}

\bibitem{Symanzik:1983dc}
K. Symanzik,
\newblock Nucl. Phys. B226 (1983) 187.

\bibitem{Symanzik:1983gh}
K. Symanzik,
\newblock Nucl. Phys. B226 (1983) 205.

\bibitem{Lepage:1996ph}
G.P. Lepage,
\newblock Nucl. Phys. Proc. Suppl. 47 (1996) 3, hep-lat/9510049.

\bibitem{Alford:1998yy}
M. Alford, T.R. Klassen and G.P. Lepage,
\newblock Phys. Rev. D58 (1998) 034503, hep-lat/9712005.

\bibitem{Hasenfratz:1994sp}
P. Hasenfratz and F. Niedermayer,
\newblock Nucl. Phys. B414 (1994) 785, hep-lat/9308004.

\bibitem{DeGrand:1995ji}
T. DeGrand, A. Hasenfratz, P. Hasenfratz and F. Niedermayer,
\newblock Nucl. Phys. B454 (1995) 587, hep-lat/9506030.

\bibitem{DeGrand:1995jk}
T. DeGrand, A. Hasenfratz, P. Hasenfratz and F. Niedermayer,
\newblock Nucl. Phys. B454 (1995) 615, hep-lat/9506031.

\bibitem{Bietenholz:1996cy}
W. Bietenholz and U.J. Wiese,
\newblock Nucl. Phys. B464 (1996) 319, hep-lat/9510026.

\bibitem{DeGrand:1998pr}
MILC, T. DeGrand,
\newblock Phys. Rev. D58 (1998) 094503, hep-lat/9802012.

\bibitem{Niedermayer:2000yx}
F. Niedermayer, P. Rufenacht and U. Wenger,
\newblock (2000), hep-lat/0007007.

\bibitem{DeGrand:cloverfat}
T. DeGrand, A. Hasenfratz and T. Kovacs,
\newblock Nucl. Phys. B547 (1999) 259.

\bibitem{Bernard:1999kc}
C. Bernard and T. DeGrand,
\newblock Nucl. Phys. Proc. Suppl. 83 (2000) 845, hep-lat/9909083.

\bibitem{Bietenholz:2000iy}
W. Bietenholz,
\newblock (2000), hep-lat/0007017.

\bibitem{DeGrand:2000tf}
MILC, T. DeGrand,
\newblock (2000), hep-lat/0007046.

\bibitem{Blum:1997uf}
T. Blum et~al.,
\newblock Phys. Rev. D55 (1997) 1133, hep-lat/9609036.

\bibitem{Lagae:1998pe}
J.F. Lagae and D.K. Sinclair,
\newblock Phys. Rev. D59 (1999) 014511, hep-lat/9806014.

\bibitem{Orginos:1999cr}
MILC, K. Orginos, D. Toussaint and R.L. Sugar,
\newblock Phys. Rev. D60 (1999) 054503, hep-lat/9903032.

\bibitem{Orginos:1998ue}
MILC, K. Orginos and D. Toussaint,
\newblock Phys. Rev. D59 (1999) 014501, hep-lat/9805009.

\bibitem{Karsch:2000ps}
F. Karsch, E. Laermann and A. Peikert,
\newblock Phys. Lett. B478 (2000) 447, hep-lat/0002003.

\bibitem{Lepage:1998vj}
G.P. Lepage,
\newblock Phys. Rev. D59 (1999) 074502, hep-lat/9809157.

\bibitem{Hasenbusch:1998yb}
M. Hasenbusch,
\newblock Phys. Rev. D59 (1999) 054505, hep-lat/9807031.

\bibitem{smear:ape}
APE, M. Albanese et~al.,
\newblock Phys. Lett. 192B (1987) 163.

\bibitem{Kluberg-Stern:1983dg}
H. Kluberg-Stern, A. Morel, O. Napoly and B. Petersson,
\newblock Nucl. Phys. B220 (1983) 447.

\bibitem{Gupta:1991mr}
R. Gupta, G. Guralnik, G.W. Kilcup and S.R. Sharpe,
\newblock Phys. Rev. D43 (1991) 2003.

\bibitem{Sommer:1994ce}
R. Sommer,
\newblock Nucl. Phys. B411 (1994) 839, hep-lat/9310022.

\bibitem{Guagnelli:1998ud}
ALPHA, M. Guagnelli, R. Sommer and H. Wittig,
\newblock Nucl. Phys. B535 (1998) 389, hep-lat/9806005.

\bibitem{Niedermayer:1997eb}
F. Niedermayer,
\newblock Nucl. Phys. Proc. Suppl. 53 (1997) 56, hep-lat/9608097.

\bibitem{Gupta:1988pf}
R. Gupta et~al.,
\newblock Phys. Rev. Lett. 61 (1988) 1996.

\bibitem{Apostolakis:1991km}
J. Apostolakis, C.F. Baillie and G.C. Fox,
\newblock Phys. Rev. D43 (1991) 2687.

\bibitem{Hasenbusch:1992tx}
M. Hasenbusch and S. Meyer,
\newblock Phys. Rev. D45 (1992) 4376.

\bibitem{Wolff:1992ze}
U. Wolff,
\newblock Phys. Lett. B284 (1992) 94, hep-lat/9205001.

\bibitem{Creutz:1987xi}
M. Creutz,
\newblock Phys. Rev. D36 (1987) 515.

\bibitem{Brown:1987rr}
F.R. Brown and T.J. Woch,
\newblock Phys. Rev. Lett. 58 (1987) 2394.

\bibitem{Decker:1990hp}
K.M. Decker and P. de~Forcrand,
\newblock Nucl. Phys. Proc. Suppl. 17 (1990) 567.

\bibitem{Gupta:1988yw}
R. Gupta, G.W. Kilcup, A. Patel, S.R. Sharpe and P. de~Forcrand,
\newblock Mod. Phys. Lett. A3 (1988) 1367.

\bibitem{Booth:1992kk}
UKQCD, S.P. Booth et~al.,
\newblock Phys. Lett. B275 (1992) 424.

\bibitem{Wolff:1992ri}
U. Wolff,
\newblock Phys. Lett. B288 (1992) 166.

\bibitem{Grady:1985fs}
M. Grady,
\newblock Phys. Rev. D32 (1985) 1496.

\bibitem{creutz_algo}
M. Creutz,
\newblock {\em Algorithms for simulating fermions}, Advanced Series on
  Directions in High Energy Physics-Vol. 11 Quantum fields on the computer (ed.
  M. Creutz) .

\bibitem{Hasenfratz:1994az}
A. Hasenfratz and T.A. DeGrand,
\newblock Phys. Rev. D49 (1994) 466, hep-lat/9304001.

\bibitem{Duncan:1998gq}
A. Duncan, E. Eichten and H. Thacker,
\newblock Phys. Rev. D59 (1999) 014505, hep-lat/9806020.

\bibitem{Duncan:1999xh}
A. Duncan, E. Eichten, R. Roskies and H. Thacker,
\newblock Phys. Rev. D60 (1999) 054505, hep-lat/9902015.

\bibitem{deForcrand:1998sv}
P. de Forcrand,
\newblock Nucl. Phys. Proc. Suppl. 73 (1999) 822.

\bibitem{Martin:1985yn}
O. Martin and S.W. Otto,
\newblock Phys. Rev. D31 (1985) 435.

\bibitem{coloprep}
A. Hasenfratz and F. Knechtli,
\newblock in preparation .

\bibitem{Gottlieb:1987mq}
S. Gottlieb, W. Liu, D. Toussaint, R.L. Renken and R.L. Sugar,
\newblock Phys. Rev. D35 (1987) 2531.

\bibitem{Duane:1987de}
S. Duane, A.D. Kennedy, B.J. Pendleton and D. Roweth,
\newblock Phys. Lett. B195 (1987) 216.

\bibitem{coloprep2}
T. DeGrand, A. Hasenfratz and F. Knechtli,
\newblock in preparation .

\bibitem{Takaishi:2000bt}
T. Takaishi and P. de~Forcrand,
\newblock (2000), hep-lat/0011003.

\bibitem{Montvay:1996ea}
I. Montvay,
\newblock Nucl. Phys. B466 (1996) 259, hep-lat/9510042.

\end{thebibliography}
\end{document}